\documentclass[%
reprint, nobalancelastpage,
superscriptaddress,
 amsmath,amssymb,
 aps,
 prx
linenumbers
]{revtex4-2}
\usepackage{comment}
\usepackage{braket}
\usepackage{graphicx}
\usepackage{dcolumn}
\usepackage{bm}
\usepackage{siunitx}
\usepackage{xcolor} 
\usepackage{tabularx}
\usepackage{dsfont}					
\usepackage{mathrsfs}	
\usepackage{float}
\usepackage{ragged2e}
\usepackage{hyperref}

\newcommand{\addAA}[1]{\textcolor{black}{#1}}
\newcommand{\addND}[1]{\textcolor{black}{#1}}
\setcitestyle{numbers,comma} 

\usepackage{lineno}
\begin{document}
\title{Nonreciprocal quantum information processing with superconducting diodes in circuit quantum electrodynamics}
\author{Nicolas Dirnegger}
\affiliation{Department of Electrical and Computer Engineering, UCLA, Los Angeles, CA, USA}
\author{Prineha Narang}
\email{prineha@ucla.edu}
\affiliation{Department of Electrical and Computer Engineering, UCLA, Los Angeles, CA, USA}
\affiliation{Division of Physical Sciences, College of Letters and Science, University of California, Los Angeles (UCLA), Los Angeles, CA, USA}
\author{Arpit Arora}
\email{arpit22@ucla.edu}
\affiliation{Department of Electrical and Computer Engineering, UCLA, Los Angeles, CA, USA}
\affiliation{Division of Physical Sciences, College of Letters and Science, University of California, Los Angeles (UCLA), Los Angeles, CA, USA}

\begin{abstract}
Introducing new components and functionalities into quantum devices is critical in advancing state-of-the-art hardware. Here, we propose superconducting diodes (SDs) as a coherent nonreciprocal element in circuit quantum electrodynamics (cQED) architectures. In particular, we use an asymmetric SQUID as an SD controlled with a flux bias -- nonreciprocal element with single control handle and on-chip modality. We spectroscopically characterize SD and show that flux bias acts cooperatively with the nonlinear diode response to induce direction-dependent resonance shifts in the transmission spectrum. We show that even with modest diode efficiency the isolation isolation ratio is sufficiently high, and scales with multiple SDs. We demonstrate the use of the SD as a coupler to realize coherent nonreciprocal qubit-qubit coupling. With a minimal two qubit system, we demonstrate nonreciprocal half-iSWAP, thereby showcasing the potential of intrinsic nonreciprocity as a tool to perform arbitrary two-qubit gates. Our work enables high-fidelity signal routing and entanglement generation in all-to-all connected microwave quantum networks, where nonreciprocity is embedded at the device level.
\end{abstract}

\maketitle
\section*{Introduction}
Nonreciprocal transfer of quantum states and correlations is crucial for scalable devices and networks~\cite{wehner2018quantum,kannan2023,barzanjeh2025nonreciprocity,almanakly2025deterministic}. In superconducting circuits, nonreciprocity is typically realized using commercially available ferrite-based components, which introduce loss and are incompatible with on-chip integration~\cite{kord2020microwave} \addAA{forming the most space-expensive component in dilution refrigerators~\cite{arnold2025all}}. Alternate superconductor based schemes have been proposed, \addAA{guided by} magnon coupling~\cite{wang2021low,chen2022nonreciprocal}, geometric designs~\cite{sliwa2015reconfigurable,muller2018passive,navarathna2023passive},  and reservoir engineering \cite{krantz2019quantum,chapman2017widely,wang2024dispersive}, to name a few. Although effective, these components have a large footprint, and active modulation adds complexity and noise. As a result, the search continues for coherent, hardware-level nonreciprocal element that can be embedded directly into superconducting circuits, enabling the design of robust and scalable quantum processors. 

The recently discovered superconducting diode (SD) effect \cite{gupta2023gate,nadeem2023superconducting, upadhyay2024microwave, ando2020observation, daido2022intrinsic, schmid2024yba} opens a new paradigm for realizing intrinsic nonreciprocal elements~\cite{labarca2024toolbox,frattini2017, frattini2018optimizing}. SDs are characterized by direction-dependent critical currents, and have been realized in bulk superconductors, Josephson junctions (JJs), and superconducting quantum interference devices (SQUIDs)~\cite{gupta2023gate}. Unlike conventional routes of nonreciprocity that rely on engineering losses \cite{ren2024nonreciprocal, metelmann2015nonreciprocal}, the superconducting diode effect arises from a fundamental asymmetry in the system response, enabled by breaking inversion ($\mathcal{I}$) and time-reversal symmetry ($\mathcal{T}$)~\cite{nadeem2023superconducting, tanaka2022theory, ghosh2024high, ma2025superconducting}. Importantly, nonreciprocity is a property of the superconducting ground state\cite{sundaresh2023diamagnetic, daido2022intrinsic}, fully captured by the free energy and the evolution of the superconducting order parameter. This functionality can be harnessed through numerous means including magnetic flux \cite{roig2024superconducting}, spin-orbit coupling \cite{mao2024universal}, spontaneously symmetry broken states \cite{anwar2023spontaneous}, SQUIDs~\cite{cuozzo2024microwave}, conformally mapped materials \cite{lyu2021superconducting} or even with chiral microwaves \cite{arora2025chiral,matsyshyn2025superconducting}. 
\addAA{SDs offer a potential route towards nonreciprocity that can be realized at fabrication-level and enable scalable modular devices with reduced complexity, e.g., more space in dilution refrigerators, hardware support for efficient lattice surgery satisfying causal information flow~\cite{thompson2018causal}.}\\

\begin{figure}
    \centering
    \includegraphics [width = \linewidth]{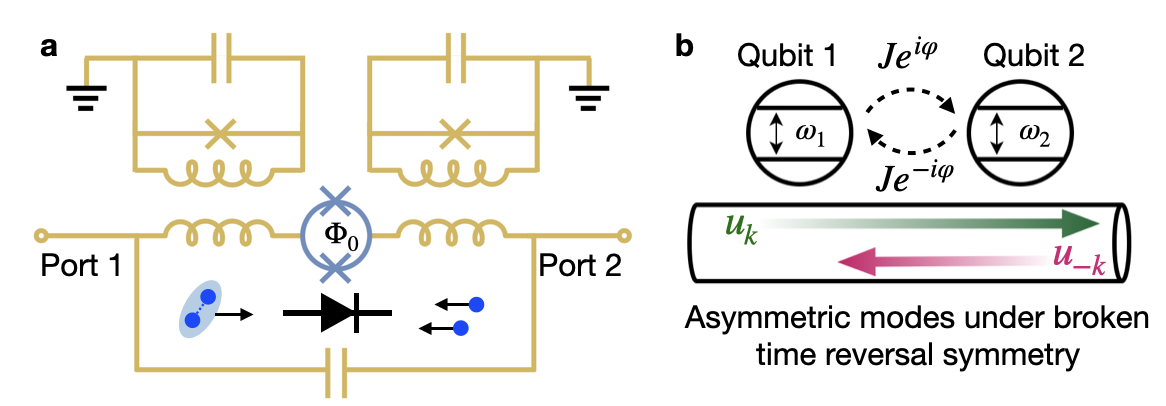}
    \caption{\justifying Superconducting diode (SD) and nonreciprocal qubit-qubit coupling: (a) Circuit of two qubits inductively coupled to a SD. Here, the SD is formed by an asymmetric SQUID (shown in blue) which is biased by external flux, $\Phi_0$. The supercurrent is allowed along the direction of forward bias (shown as bound Cooper pairs), and is blocked along the direction of reverse bias (shown as broken Cooper pairs). (c) Schematic of nonreciprocal coupling due to SD. The broken time reversal ($\mathcal{T}$)-symmetry necessary to induce a diode response generically makes the coupling between qubits to be complex, $J_{12} = Je^{i\phi_{12}}$ with non-local phase $\phi_{12}=\varphi$ is determined by nonreciprocal response by asymmetric mode propagation in SD.}
    \label{fig:Setup}
\end{figure}

In this work,
we consider an asymmetric SQUID as an SD -- \addAA{a device compatible with state-of-the-art fabrication standards --} with a flux-switching bias which cooperates with nonlinear inductance to enable the nonreciprocal response, as shown in Fig.~\ref{fig:Setup}a,b. We calculate the transmission spectrum of the SD in a circuit quantum electrodynamics (cQED) setup, featuring an asymmetric mode structure with direction-dependent resonance shifts. 
\addND{Furthermore, we demonstrate the SD's ability as a scalable alternative to state-of-the-art isolators and circulators by simulating the isolation ratio with single and multiple diodes. }We then obtain coherent nonreciprocal qubit-qubit interactions determined by complex coupling mediated by the SD. We analyze the corresponding qubit population dynamics, and implement nonreciprocal entangling~\cite{sung2021realization} two-qubit gates using this architecture. 
These results establish the SD as a resource of nonreciprocity embedded at the device level and a scalable building block for directional quantum control.

\section*{Results}

\textbf{Superconducting diode, spectroscopic characterization and readout potential} \\

The SD effect is a magnetoelectric nonlinear response in superconductors and superconducting devices. Microscopically, the diode behavior emerges from a shift in Cooper pairs center of mass, fundamentally realized by simultaneous breaking of $\mathcal{T}$ and $\mathcal{I}$ symmetries. \addAA{In JJ and SQUID based devices, SD is phenomenologically captured as $I(\Phi) = I_a\sin(\Phi) + I_b\sin(2\Phi-\delta)$~\cite{gupta2023gate, ingla2025efficient} where $I_{a,b}$ are the magnitudes of fundamental and higher harmonics, respectively, and $\delta$ is the relative phase; $\delta\neq 0$ for broken $\mathcal{T},\mathcal{I}$.} 
The SD is characterized by its diode efficiency~\cite{nadeem2023superconducting, schmid2024yba}

\begin{equation}
    \eta = \left|\frac{I_c^+ - |I_c^-|}{I_c^+ + |I_c^-|}\right|
\end{equation}

where $I^\pm$ is the current in forward/backward direction, and $I_c^\pm$ is the corresponding critical current. Recent experiments have demonstrated SD behavior in a variety of platforms, including noncentrosymmetric thin films such as Nb$_3$Br$_8$ \cite{wu2022field}, proximitized semiconductors with spin-orbit coupling (e.g., InAs/Al Josephson junctions) \cite{ciaccia2023gate,bhowmik2025optimizing,leroux2022nonreciprocal}, and asymmetric SQUID~\cite{li2025field, fominov2022asymmetric}. Additionally, magnetic-field-free diodes have been created from graphene-based devices \cite{lin2022zero} and high temperature effect \addAA{($\sim 80$ K)} has been reported for anomalous cuprate junctions \cite{kokkeler2022field}. 

Here, we consider an SD formed by an asymmetric SQUID, where the diode response is controlled by external bias flux ($\Phi_{b}$, see Fig.~\ref{fig:Setup})~\cite{souto2022josephson}. The asymmetric SQUID is characterized by the potential $U(\Phi) = U_1(\Phi) + U_2(\Phi - \Phi_b)$, where $U_{1,2}(\Phi) = -\Delta[1-\tau_{1,2}\sin^2(\Phi/2)]^{1/2}$ is the Josephson potential of JJs in the SQUID, $\tau_{1,2}$ is the junction transmission, and $\Delta$ is the superconducting gap. The diode respose arises from $\mathcal{T},~\mathcal{I}$-broken quantum interference between high and low transmission Andreev bound states in the SQUID: $\mathcal{T}$ is broken by $\Phi_b\neq n\pi$ ($n\in$ integer) and $\mathcal{I}$ is broken by $\tau_1\neq \tau_2$. 
We summarize the fundamentals of diode response in the SQUID in Supplementary Information (SI) section S1.

We define the notion of directionality and bias of the SD: sign of flux, $\Phi_b$, sets the bias of diode and applied current, $I_a = I^\pm$, sets the signal directionality. In other words, $\Phi_b = \pm|\Phi_b|$ sets the diode in forward (reverse) bias allowing $I_a=I_\pm$ to pass and blocking $I_a = I_\mp$. For cQED modalities this translates to nonreciprocity in the kinetic inductance, $L_\pm = L_0\left[ 1+(I_a/I_c^\pm)^2\right]$ where $L_0$ is the characteristic circuit inductance. This can be directly seen in spectroscopic signatures when the SD is shunted to a capacitor, $C$ with a transmission spectrum peak at $\omega_r^\pm = 1/\sqrt{L_\pm C}$. 

\begin{figure*}
    \centering
    \includegraphics[width = \linewidth]{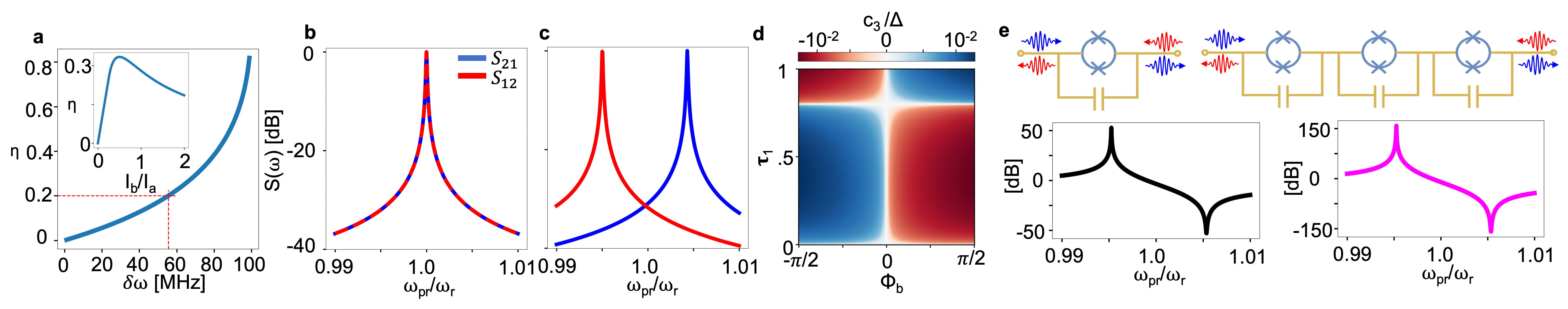}
    \caption{\justifying Spectroscopic characterization of superconducting diode on a two-port network: (a) Variation of $\delta\omega$ in variation with $\eta$ for $I_a = 30$ nA, $I_c^- = 150$ nA and $\omega_r = 5$ GHz. Inset shows variation of $\eta$ with $I_b/I_a$ for $\delta = \pi/2$. (b) Transmission spectrum at fixed off-resonant pump frequency $\omega_p /\omega_r= 0.99$, normalized by resonator frequency $\omega_r = 5$ GHz, shows scattering matrix $S(\omega)$ with respect to probe frequency $\omega_{pr}/\omega_r $ for $\delta\omega = 0$. (c) Transmission spectrum with  $\delta\omega = 50$ MHz showing clear asymmetric shift between $S_{21}$ and $S_{12}$. (d) Third-order Josephson nonlinearity, $c_3/\Delta$ with , for the asymmetric SQUID versus flux $\Phi_b$, and transmission, $\tau_1$ ($\tau_2 = 0.8$). (e) Isolation ratio for single asymmetric SQUID Diode (black) and three in series asymmetric SQUID diodes (red). The isolation ratio scales linearly with the amount of superconducting diodes and reaches isolation ratios beyond 20dB. For the three SD setup, we take into account the transmission line effects connecting all three in the analysis. (f) Schematic depiction of single (upper) and three superconducting diodes (lower) with applied flux $\Phi_0$.  For all simulations the following parameter were used $\omega_p / \omega_r = 0.99$, Kerr nonlinearity $\Lambda/\omega_r = 10^{-7}$, two photon loss $\kappa/\omega_r = 10^{-4}$.}
    \label{fig:Spectroscopy}
\end{figure*}

The corresponding frequency shift can be obtained using the lumped element \addND{approach} where the measured frequency is $\omega^\pm_r=\omega_r\sqrt{L_0/L_\pm}$. We have defined $\omega_r = 1/\sqrt{L_0C}$ as the characteristic frequency of the circuit. The shift in resonance is $\delta \omega_\pm/\omega_r\approx -0.5(I_a/I_c^{\pm})^2$. For diode response, we have $I_c^+ \neq I_c^-$ leading to $L_+\neq L_-$. For $\eta\approx 20\%$, we have $I_c^+/I_c^- = 3/2$ which corresponds to $L_K^- - L_K^+ = 5L_0I_a^2/[9(I_c^-)^2]$. Then, we estimate $\delta\omega = \delta\omega_- - \delta\omega_+$ using standard cQED experimental inputs $\omega_0=5$ GHz, $I_a = 30$ nA and $I_c^- = 150$ nA which corresponds to $\delta\omega\approx 55$ MHz. In Fig.~\ref{fig:Spectroscopy}a, we show the variation of $\delta\omega$ in variation with $\eta$; inset shows variation of $\eta$ with $I_b/I_a$ with $\delta = \pi/2$.

The nonreciprocal behavior can be tracked via transmission spectrum.
We follow standard recipe of input-output theory (also see SI section S2), according to which $S_{ji}(\omega) = Z_0/(Z_0 + Z_{ji}(\omega))$ where $S_{ji}(\omega)$ is the transmission, $Z_{ji}(\omega)$ is the impedance response of the circuit and $Z_0$ is the matched impedance between the two ports. For resistance $R$ in the circuit $Z_{ji}(\omega) = R + i [\omega L_{ji} - (\omega C)^{-1}] $, and for SD $L_{ji} \neq L_{ij}$. Consequently, we obtain forward $S_+(\omega) = S_{21}(\omega)$ and backward transmission $S_-(\omega) = S_{12}(\omega)$ as 
\begin{equation}
\label{eq:transmission}
    S_{\pm}(\omega) = \frac{\mathcal{A}_{\pm}}{(\omega - \omega_{\pm})^2 + \kappa^2_{\pm}}
\end{equation}
where $\mathcal{A}_{\pm} = (Z_0/2L_{\pm})^2$ and $\kappa^\pm = Z_0+R/2L_{\pm}$. Crucially, with $S_+(\omega) \neq S_-(\omega)$ Eq.~(\ref{eq:transmission}) serves as the central relation to the benchmark an SD. In Fig.~\ref{fig:Spectroscopy}b,c, we show the transmission spectrum calculated using Heisenberg-Langevin equations at mean-field level (see SI section S3 for details) for $\delta\omega = 0$ and 50 MHz, respectively. We account for Kerr nonlinearity, drive and \addND{dissipation} to show that the SD induced nonreciprocity is robust. 

Next, as a proof of concept, we elucidate the microscopic origin of SD nonreciprocity using the canonical Hamiltonian for SD shunted to capacitor: $H_r = 4E_c N + U(\Phi)$ where $E_c$ is the charging energy and $U(\Phi)$ is the Josephson energy potential. We Taylor expand $U(\Phi) = \sum_n\frac{c_n}{n!} (\Phi-\Phi_{\rm min})^n$ around minima $\Phi_{\rm min}$ obtained self-consistently by solving $c_1 = \partial_{\Phi}U(\Phi) = 0$~\cite{frattini2018optimizing}. The diode nonreciprocity emerges as a third order nonlinearity, $c_3(\Phi_b) \Phi^3/6$ with $c_3 = \partial_{\Phi}^3U(\Phi)|_{\Phi_{\rm min}}$. Note that $c_3(\Phi_b)\neq 0$ explicitly breaks $\mathcal{T}$ as $U(-\Phi)\neq U(\Phi)$, and it is odd with respect to $\Phi_b$ vanishing identically at $\Phi_b = 0$, see Fig.~\ref{fig:Spectroscopy}c. This contribution has been known as a source of three-wave mixing, seen also in Josephson ring modulators ~\cite{abdo2013nondegenerate}, and recently discussed in Ref.~\cite{schrade2024dissipationless} in context of minimizing Kerr-nonlinearity in SDs, akin to SNAIL parametric amplifiers ~\cite{frattini2018optimizing, sivak2019kerr}. The combination of $\Phi_b,c_3(\Phi_b)\neq 0$ leads to bias and direction controlled resonance shifts which can be obtained by transforming to bosonic operators. We define
\begin{equation}
\label{eq:bosonoperator}
    \Phi(x) = \sum_k \Phi_{\rm zpf} \left[\psi_k(x) a_k + \psi^*_k(x)a_k^\dagger\right]
\end{equation}

where $a_k$ is the bosonic operator, and $\psi_k(x) = u_k e^{ikx}$ is the mode function between the ports; $\Phi_{\rm zpf}$ is the flux at zero point fluctuation. Within mean-field treatment we write $a_k = \alpha_k +d_k$ where $\langle a_k\rangle = \Phi_b\alpha_k$ is the mean-field expectation value satisfying $\langle\Phi\rangle = \Phi_b$ and $\langle d_k\rangle = 0$. The third order nonlinearity in Josephson potential at leading order in $\Phi_b$ gives $c_3(\Phi_b)\Phi_b \sum_{k}d_k^\dagger d_k$ where the bilinear form of the bosonic operators beyond three-wave mixing together with $u_{-k}^*\neq u_k$ under broken-$\mathcal{T}$ results in direction dependent resonance frequency shift, $\delta\omega=\omega_{k} - \omega_{-k}$ such that  

\begin{equation}
\label{eq:frshift}
    \delta\omega = {\rm sgn}(\Phi_b) |c_3(\Phi_b)\Phi_b \mathcal{F}_k(\Phi_b)|
\end{equation}
where $\mathcal{F}_k(\Phi_b)$ captures the geometry dependent factor odd in $\Phi_b$ due to mode asymmetry, $u_k \neq u_{-k}^*$, see SI section S4 for details. Therefore, we have a resonator with flux-controlled resonance shifts when the system responds with $c_3(\Phi_b)\neq0$ under broken $\mathcal{T}$ and $\mathcal{I}$.

\addND{Due to the SD's internal mechanism to shift the resonant frequency via applying the flux bias, we demonstrate the devices ability to isolate signals and block signals from the opposing direction. In Fig. \ref{fig:Spectroscopy}e we show the isolation ratio defined here as $S_{21}/S_{12}$ for $\delta \omega =50$ MHz ($\eta \approx 20\%$), where a single SD is able to surpass the 20 dB benchmark for isolation ratio, although for a narrow bandwidth up to 100kHz, which can be further enhanced by connecting multiple SD resonators in series, see Fig.~\ref{fig:Spectroscopy}e where we use three SDs. The inherent ability of the SD to isolate signals and protect them make it a suitable candidate for applications such as readout. However, further analysis is required to determine the SD's feasibility on a hardware level in such a context.}\\

\textbf{Nonreciprocal qubit-qubit coupling by SD}\\

Having established the characterization of the nonreciprocal SD through Fig.~\ref{fig:Spectroscopy} and Eq.~(\ref{eq:transmission}), we use it to induce nonreciprocal coupling between qubits. We consider a 2-qubit ($Q_1,Q_2$) system for brevity but our approach is valid for $n$-qubits connected through a single resonator \cite{kafri2017tunable}. The effective Hamiltonian for the 2-qubit system in leading order in qubit operators can be generically expressed as  

\begin{equation}
\label{eq:qubitH}
    \mathcal{H} = \sum_i \frac{\omega_i}{2} \sigma_z^{(i)}+ \sum_{i\neq j}J_{ij} (\sigma_-^{(i)} \sigma_+^{(j)} + {\rm h.c.})
\end{equation}
where $\sigma_\pm^{(i)} = \sigma_x^{(i)}\pm\sigma_y^{(i)}$ and  $\sigma_{x,y,z}^{(i)}$ is Pauli operator for qubit at port $i\in\{1,2\}$. Here, $\omega_{i}$ is the excitation energy of qubit $i$ and $J_{ij}$ is the interaction between nearest neighbor qubits $i$ and $j$. Importantly, $J_{ij}$ captures the nonreciprocity carried by the SD resonator, and can be expressed with a complex phase
\begin{equation}
\label{eq:complexJ}
    J_{ij} = |J|e^{i\phi_{ij}}.
\end{equation}
where $J$ is the strength of the coupling and $\phi_{ij}$ is a non-local phase governing the nonreciprocal interactions between qubits located at different ports of the resonator. 

\begin{figure*}
    \centering
    \includegraphics[width=\linewidth]{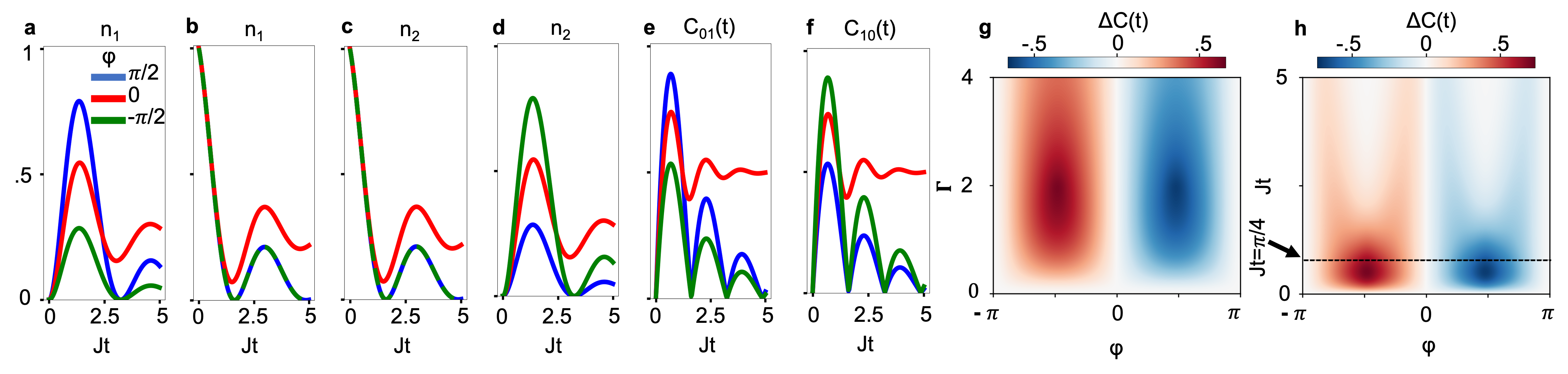}
    \caption{\justifying Coherent control of population dynamics via effective qubit-qubit coupling through SD and dynamic evolution of concurrence in two qubit system. (a-d) The population dynamics of qubits $n_1(t),n_2(t)$ at $\varphi \in [-\pi/2,0,\pi/2]$ (green, red and blue lines). Panels a,b show $n_1(t),n_2(t)$, respectively with $n_1(t=0),n_2(t=0)=1$, and the system initialization is reversed in panels c,d.
    (e,f) Concurrence $C_{01}$ and $C_{10}$ plotted against $J t$ for different $\varphi \in [-\pi/2, 0, \pi/2]$.  
    For $\varphi= 0$ (red) the concurrence $C_{01}$ and $C_{10}$ are identical. For $\varphi = \pm \pi/2$ (blue for $C_{01}$ and green for $C_{10}$), we observe that the concurrence grows close to 1.0 signaling maximal entanglement at time scale of half-iSWAP facilitated by SD nonreciprocity. 
    (g) Heatmap for entanglement transfer contrast, $\Delta C(t)$ at $t=\pi/4J$ as a function of $\varphi$ and $\Gamma$. $\Delta C(t=\pi/4J)$ is maximum at $\Gamma/J = 2$ and $\varphi = \pm\pi/2$, and significant nonreciprocity sustains throughout the phase space. (h) $\Delta C(t)$ in variation with $Jt$ and $\varphi$ showing maximum entanglement contrast at $\varphi = \pm \pi/2$ at the scale of time of half-iSWAP. 
    All plots were made with following parameters $\gamma_{1} = 0$, $J=1$, $\Gamma =  0.5$, unless mentioned otherwise.}
    \label{fig:Pop}
\end{figure*}

Phenomenologically, Eq  .~(\ref{eq:complexJ}) captures fundamental nonreciprocity in any system breaking $\mathcal{T}$ symmetry~\cite{koch2010time,anderson2016engineering,labarca2024toolbox,ren2024nonreciprocal}. To track its microscopic origin within the diode response of SD, we calculate the qubit-qubit coupling 
\begin{equation}
\label{eq:coupling-green}
    J_{ij} = g^2\sum_k  \mathcal{G}_{ij,k}(\omega) 
\end{equation}
by evaluating the causal Green's function between the ports, $\mathcal{G}_{ij,k} = \psi_k(x_i) \chi_k(\omega) \psi_k(x_j)^*$ with $\chi_k(\omega)=[(\omega - \omega_k) + i\kappa]^{-1}$; $\kappa$ is two photon decay. Without loss of generality we have assumed qubits couple to resonator identically with coupling strength $g$, and mode amplitudes are defined as in Eq.~(\ref{eq:bosonoperator}). We decompose the Green's function into forward and backward traversing modes such that $\mathcal{G}_{ij,k} = \mathcal{G}_{ij,+k} +\mathcal{G}_{ij,-k} $ and restrict the summation in Eq.~(\ref{eq:coupling-green}) to $k>0$. This allows us to extract the reciprocal and nonreciprocal counterparts of the coupling such that $J_{ij} = J_{ij}^{\rm r} + i J_{ij}^{\rm nr}$ reproducing Eq.~(\ref{eq:complexJ}) with the non-local phase
\begin{equation}
\label{eq:phase}
    \phi_{ij} = \tan^{-1}\left(\frac{J_{ij}^{\rm nr}}{J_{ij}^{\rm r}}\right)
\end{equation}
characterized by the non-reciprocal response of the system
\begin{equation}
\label{eq:Jnr}
    J_{ij}^{\rm nr} = g^2\sum_{k>0} |\chi_{k}(\omega)|^2 \sin[k(x_j-x_i)] \delta\omega.
\end{equation}
Importantly, we have obtained SD determined nonreciprocal qubit-qubit coupling which is a consequence of fundamental device response and controlled by $\Phi_b$ and direction of signal, see Eq.~(\ref{eq:frshift}) and Eq.~(\ref{eq:Jnr}) -- $J_{ij}^{\rm nr}$ flips sign with $\Phi_b$ and $(x_i \leftrightarrow x_j)$. Strikingly, Eq.~(\ref{eq:complexJ},\ref{eq:phase}) match up with general nonreciprocal coupling obtained in Ref.~\cite{labarca2024toolbox} for a general $\mathcal{T}$-broken quantum model. 

In what follows, we use $\phi_{ij} = \varphi$ as a parameter to demonstrate nonreciprocal qubit dynamics for Eq.~(\ref{eq:qubitH}). 
Before proceeding to demonstrate the nonreciprocity, we emphasize that $\phi_{ij} \neq 0$ is determined by the behavior of SD, and allows control of nonreciprocity of the system by device response, unlike existing paradigms which rely on dissipation controlled  non-Hermitian dynamics of the system. However, given the passive nature of nonreciprocity, the presence of drive or dissipation is necessary to translate SD characteristics to system dynamics. Nevertheless, as we show below the control handle is completely determined by control over SD.

We model the dynamics of our minimal system consisting of SD resonator coupled to two qubits as an open quantum system and following standard recipe of Lindblad master equation track qubit population through time evolution of $\sigma_z$
\begin{align} \label{eq:EOM}
    \dot{\sigma}_z^{(i)} 
    &= -\gamma_{1,i} \bigl(\sigma_z^{(i)}+1\bigr) 
    - 2i \sum_{j\neq i} J_{ij} \bigl( \sigma_+^{(i)} \sigma_-^{(j)} - \sigma_+^{(j)} \sigma_-^{(i)} \bigr) \nonumber\\
    &\quad - \sum_{j\neq i} \Gamma \bigl( \sigma_+^{(i)} \sigma_-^{(j)} + \sigma_+^{(j)} \sigma_-^{(i)} \bigr),
\end{align}
where we have accounted for qubit relaxation ($\gamma_{1,i}$) and collective cross-qubit decay ($\Gamma$). The qubit population is obtained as $n_i(t) = \langle\sigma_z^{(i)}(t)\rangle$. The qubit initialization is described in terms of initial condition of Eq.~(\ref{eq:EOM}), $n_i(t=0)$.

In Fig.~\ref{fig:Pop}a,b, we show $n_{1,2}(t)$ with $n_{1,2}(t=0) = 0,1$, and in Fig.~\ref{fig:Pop}c,d we show $n_{1,2}(t)$ with flipped system initialization, $n_{1,2}(t=0) = 1,0$. 
When $\varphi = 0$ (red curves), the population dynamics are reciprocal: regardless of whether qubit-1 or qubit-2 is initially excited, the subsequent time evolution is identical, reflecting symmetric coupling between the two modes. In contrast, for $\varphi = \pm \pi/2$ (blue and green curves), the dynamics become nonreciprocal. Depending on the sign of $\varphi$, excitation transfer is either enhanced or suppressed in one direction as compared to the other, leading to a pronounced directional asymmetry. This phase-controlled nonreciprocal population transfer effectively mimics isolation: the system favors energy flow in one direction while inhibiting it in the reverse. 
At long times, the exponential decay of all curves reflects intrinsic dissipation, which sets the overall relaxation timescale.\\

\textbf{Directional half-iSWAP gate}\\

\addAA{To leverage the nonreciprocal dynamics in the context of quantum information processing, we demonstrate a directional transverse two-qubit gate defined by the unitary operator:}
\begin{equation}\label{eq:unitary}
    U = \begin{pmatrix}
        1 & 0 & 0 & 0 \\
        0 & \cos(Jt) & -ie^{-i\varphi}\sin(Jt) & 0\\
        0 & -i e^{i\varphi}\sin(Jt) & \cos(Jt) & 0 \\
        0 & 0 & 0 & 1
    \end{pmatrix}
\end{equation}
\addAA{which follows from time evolution of Eq.~\ref{eq:qubitH}, where $\theta = Jt$ represents the swap angle. The operator generates maximally entangled states at $t=\pi/4J$ indicative of the half-iSWAP gate.} We project onto the one-excitation subspace $\{|10\rangle,|01\rangle\}$, and starting with $|\Psi(t=0)\rangle = |01\rangle$, the corresponding entangled state at the time of half-iSWAP, $t=\pi/4J$, is then obtained as 
\begin{equation}
\label{eq:bell-state-diode}
   \left| \Psi(t=\pi/4J) \right\rangle= \frac{1}{\sqrt{2}}\left(|01\rangle + i e^{i\varphi}|10\rangle\right),
\end{equation}
and the standard Bell states are obtained at $\varphi=\pm\pi/2$. Such tunability of symmetric and asymmetric combination of wavefunctions is reminiscent of tunable couplers \cite{yan2018tunable,ding2023high}. However, in contrast to tunable couplers the entanglement generation, as we show below, is nonreciprocal. \addAA{Additionally, $\varphi\neq 0$ leads to new channels in two-qubit gates which though do not fall under Clifford gates but can be utilized for simulations \cite{foxen2020demonstrating}}. 
\addAA{The distinction of $e^{i\varphi}$ must be noted against single qubit rotation as $e^{i\varphi}$ acts only on the qubit to which population is transferred. A unitary operator similar to that of Eq.~(\ref{eq:unitary}) was obtained by driving tunable coupler with both dc-bias and high-speed flux line with phase being reciprocal determined by parametric drive~\cite{mckay2016universal,roth2017analysis,ganzhorn2019gate}, in contrast to the SD determined phase $\varphi$.}

To highlight the phase-controlled directionality of entanglement transfer, we evaluate the concurrence $C(t)$ following Ref.~\cite{wootters1998entanglement}, which quantifies bipartite entanglement on a scale from 0 (separable) to 1 (maximally entangled). Figures~\ref{fig:Pop}e,f show the time evolution of $C_{01}(t)$ and $C_{10}(t)$ corresponding to system initialization at $\ket{\Psi(t=0)} = \ket{01}$ and $\ket{\Psi(t=0)} = \ket{10}$, respectively. For $\varphi=0$, the concurrence dynamics are symmetric and yield identical entanglement generation for both initial conditions. In contrast, for $\varphi\neq 0$, the dynamics become strongly asymmetric: for $\varphi = +\pi/2$ the entanglement generated when qubit 1 is initially excited is significantly enhanced with maximal entanglement achieved at $t=\pi/4J$. For $\varphi = -\pi/2$ the situation is reversed, with maximal entanglement occurring when qubit 2 is initially excited. In Figs.~\ref{fig:Pop}g, we show the entanglement transfer contrast $\Delta C(t) = C_{01} (t) - C_{10} (t)$
as a function of $\varphi$ and $\Gamma$ at $t=\pi/4J$. Previous studies on two-port nonreciprocal qubit coupling relied on dissipation engineering to find sweet-spot in parameter space at $2J = \Gamma$~\cite{ren2024nonreciprocal,nefedkin2024theory}. Strikingly, in contrast we find that though presence of dissipation is necessary given the passive of nature of SD nonreciprocity but high degree of isolation is achievable even away from $2J = \Gamma$. This can be seen as wide spread blue (for $\varphi >0$) and red (for $\varphi <0$) regions where $|01\rangle$ and $|10\rangle$ initialized states dominate, respectively. 
Finally, in Fig.~\ref{fig:Pop}h, we show $\Delta C(t)$ in variation with $\varphi$ and $t$ at $\Gamma = J$, and note maximal isolation around $\varphi = \pm\pi/2$ and $t=\pi/4J$.
The transient character of peaks indicates dominant coherent interactions driving nonreciprocal entanglement dynamics at scale of half-iSWAP time. \\

\addAA{Next, we demonstrate a verifiable advantage of SD. We perform two‐qubit state tomography at \(t = \pi/4J\). 
We stimulate the standard 16‐basis measurement protocol \cite{James2001,Paris2004,Lvovsky2009, Mahler2013,Ferrie2014,DiCarlo2009,Egger2014,Bialczak2010,Gambetta2011,Eichler2011,Neeley2010,Kamal2011,Abdo2013,Bernier2017,Ranzani2015} (combinations of \(\{I,X,Y,Z\}\) on each qubit) using linear reconstruction. While our starting point is the numerically evaluated denisty matrix, $\rho(t)$, in presence of collective cross-decay ($\Gamma$), the tomographic representation provides a clear benchmark to probe SD advatange in directional signal routing, (SI section S6 for more details). We supplement the tomographic visualization with CHSH parameter, determined via the Horodecki criterion~\cite{horodecki2009quantum}, to show that the half-iSWAP indeed generates nonreciprocal non-deterministic entanglement.}

\begin{figure*}
    \centering
    \includegraphics[width=\linewidth]{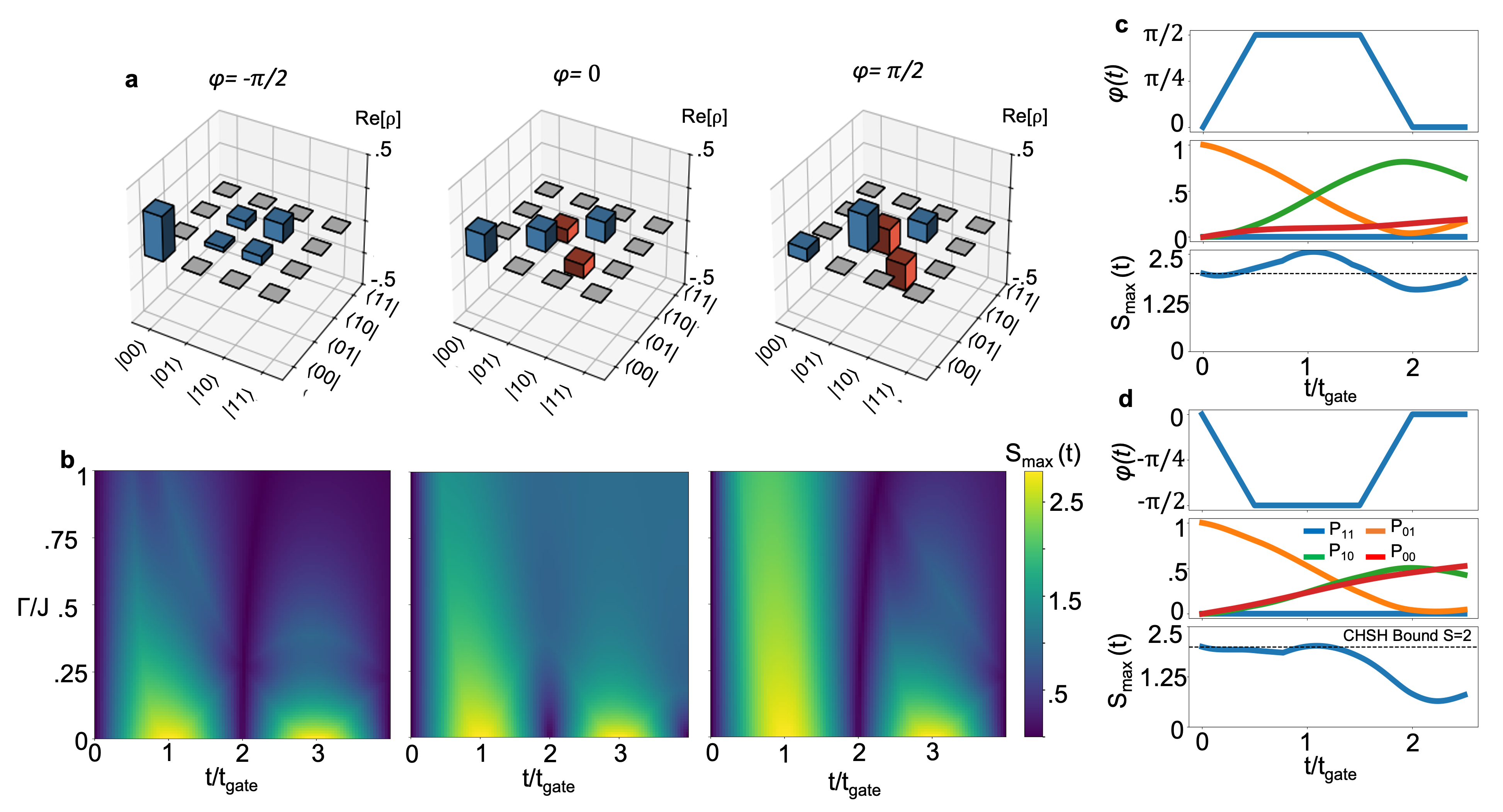}
    \caption{Advantage of SD in two-qubit entangling gate. (a) Quantum State Tomography for $\varphi = -\pi/2,0, \pi/2$ showcasing reciprocal vs. nonreciprocal Bell state creation depending under dissipation. For $\varphi =0$ we obtain a fidelity of 60\% at $\Gamma =1$, while for the nonreciprocal cases we obtain for $\varphi = \pi/2$ a fidelity of 80\% and $\varphi=-\pi/2$ a fidelity of 30\% respectively. (b) CHSH verification to show quantum advantage of nonreciprocal half-iSWAP gate. We use the Horodecki criterion \cite{horodecki2009quantum} to determine the maximum possible CHSH value $S_{\rm max}$. We plot over dissipation $\Gamma$ and normalized time $t/t_{\rm gate}$ the changes in $S_{\rm max}(t)$. $S_{\rm max}>2$ verifies the non-determinisctic entanglement for SD mediated entanglement. Importantly, for $\varphi=\pm\pi/2$, the values $S_{max}$ exceeds (lags behind) the reciprocal case. (c,d) Pulsed gate operations modeled as pulse variation of $\varphi(t)$ for $\varphi = \pi/2$ (c) and $\varphi = -\pi/2$ (d). We plot the pulse control of a trapezoid pulse in the top part. Middle panel shows population dynamics under pulse control. Bottom panel showcases $S_{max}(t)$ for $t/t_{gate}$. Black dotted line shows CHSH bound for classical system where $S_{max} = 2$. All plots were made with following parameters $\Gamma=1$, $\gamma_{1} = \gamma_{2} = \gamma_{\phi_1} = \gamma_{\phi_2} = 0$, $J=1$ unless mentioned otherwise.}
    \label{fig:BellStateTomo}
\end{figure*}

\addAA{In Fig.~\ref{fig:BellStateTomo}a, we show the two-qubit quantum state tomography for $\varphi = -\pi/2, 0, \pi/2$ (left to right). The system was initialized in $|\Psi(t=0)\rangle = |01\rangle$, and with $\Gamma = 1$ we calculate the density matrix, $\rho (t=t_{\rm gate} = \pi/4J)$. For $\varphi = 0$, a Bell state is formed with a fidelity of $60\%$, and the fidelity goes up ($80\%$) when the SD is turned on at $\varphi = \pi/2$ proving the advtange of SD mediated qubit coupling; fidelity decreases ($30\%$) for $\varphi = -\pi/2$ proving the reduced back-propogation error. This is directly reflected in the maximum CHSH parameter value, $S_{\rm max}$ for the two-qubit system. As shown in Fig.~\ref{fig:BellStateTomo}b, for $\varphi = -\pi/2$, $S_{\rm max}$ decreases rapidly wih increase in $\Gamma$, as compared to $\varphi = 0$. On the contrary, for $\varphi = \pi/2$, $S_{\rm max}$ remains above the CHSH bound of $S_{\rm max} = 2$ even for significantly high $\Gamma$. Finally, we model $\varphi$ as a signal through the flux cable (SI section S5 for details) and track the entanglement generation. In Fig.~\ref{fig:BellStateTomo}c,d we consider $\varphi(t)$ as a trapezoidal pulse (top panels) and track $S_{\rm max}(t)$ (bottom panles) along with two-qubit population $P_{00},P_{01},P_{10},P_{11}$ (middle panels) obtained from diag[$\rho(t)$]. For $\varphi = -\pi/2$ no entanglement occurs, also see Fig.~\ref{fig:BellStateTomo}a, and the state initialized with $P_{01}=1$ losses it population under decay as $P_{00}$ increases significantly. This is also also reflected in $S_{\rm max}<2$ for all $t$. On the other hand, for $\varphi=\pi/2$, the increase in $P_{00}$ is much slower and $S_{\rm max}>2$ even during parts of ramp-up and ramp-down times of the pulse.}

\section*{Discussion}
\addAA{We return to $U$ in Eq.~(\ref{eq:unitary})  and comment on the utility of SD in context of fSim gate~\cite{foxen2020demonstrating}, which can perform an arbitrary two-qubit gate within the excitation-preserving subspace, and fermionic simulations~\cite{ganzhorn2019gate}. Eq.~(\ref{eq:unitary}) along with $ZZ$-interactions represents the generalized nonreciprocal fSim gate which maps a fermionic system with nonreciprocal hopping. The SD mediated interaction realizes a native Peierls-substituted fermionic Hamiltonian of the form $t_{ij} e^{i\phi_{ij}} + {\rm h.c.}$ originating from intrinsic time-reversal symmetry breaking and corresponds to a gauge-invariant synthetic vector potential. On extended lattices, this enables the engineering of nonzero plaquette flux, programmable Chern bands, and chiral many-body dynamics without digital Trotterization overhead~\cite{heyl2019quantum}. The resulting platform, to be studied in future works, naturally accesses interacting fermionic models with nontrivial magnetic (e.g., Dzyaloshinskii-Moriya) interactions, directional transport, providing a hardware-native route to topological and chiral quantum matter simulation in superconducting circuits.}

Altogether, we proposed and demonstrated the SD as a coherent and passive nonreciprocal element featuring a single control handle capable of directional quantum state transfer determined by its intrinsic device response. This functionality was most transparently demonstrated with a minimal two-qubit architecture, and can be naturally extended to multi-qubit systems and distributed interconnects ~\cite{kannan2020waveguide,Ranzani2015}. We considered an asymmetric SQUID realization of the SD ~\cite{he2023supercurrent, golod2022demonstration}, which can be readily integrated with state-of-the-art circuit-QED hardware~\cite{wang2024dispersive}. The spectroscopic characterization and demonstration of nonreciprocal two-qubit gates benchmarks developed here provide practical routes for implementing these functionalities in near-term experiments. 
\addAA{Our work lays groundwork for easy transition from the theoretical framework presented here to scalable hardware, a clear developmental path that can be realized by making asymmetric SQUIDS with state-of-the-art Al/Al$_2$O$_3$ standards and benchmarking them against signatures presented in this work.} 

Looking forward, our demonstration of intrinsic superconducting nonreciprocity opens a clear path to on-chip signal routing and isolation without bulky circulators ~\cite{Kamal2011, barzanjeh2017mechanical, zhang2021magnetic}.  By embedding SD resonators directly into cQED chips, one can dramatically reduce cryogenic wiring and footprint.  The SD’s single handle “on/off” knob also promises hardware-level multiplexing and demultiplexing, cutting resource overhead by orders of magnitude. Tunable nonreciprocal phases afford a versatile tool to realize directional two- and multi-qubit gates (e.g. cascaded iSWAP chains) ~\cite{sliwa2015reconfigurable, kerckhoff2015chip}.
Exploring these gates within small repetition or surface-code modules may yield improved fault-tolerance thresholds through reduced back-action~\cite{ofek2016extending}.
Moving beyond proof-of-principle, key next steps include device optimization, detailed coherence studies, and full-chip simulations of SD-interconnected qubit lattices ~\cite{arute2019quantum}. Ultimately, embedding nonreciprocal superconducting elements at the device level could transform how we build modular processors.

\section*{Data availability}
Data that support the findings of this study are available from the corresponding authors upon reasonable request. These can be reproduced using custom codes (see the Code availability statement).

\section*{Code availability}
The codes used in this study are available from the corresponding authors upon reasonable request.

\bibliographystyle{apsrev4-2}
\bibliography{references}

\begin{acknowledgements}
We acknowledge helpful discussions with William Oliver, Joel Wang, Aziza Almanakly, David Pahl and Murat Can Sarihan for providing device level insights and useful comments on the manuscript. We also thank Lukas Baker, Othmane Benhayoune-Khadraoui, and William Munizzi for useful discussions. This work was supported by the Defense Advanced Research Projects Agency (DARPA), Quantum Science Center (a National Quantum Information Science Center of the U.S. Department of Energy), Gordon and Betty Moore Foundation (Grant Numbers GBMF8048 and GBMF12976 ), and the John Simon Guggenheim Memorial Foundation (Guggenheim Fellowship).
\end{acknowledgements}

\section*{Author contributions}
A.A and P.N. conceived the project. N.D. and A.A. performed the calculations and analysis. All authors wrote the manuscript.

\section*{Competing interests}
The authors declare no competing interests.

\appendix
\onecolumngrid
\newpage
\pagebreak
\widetext
\setcounter{equation}{0}
\setcounter{figure}{0}
\setcounter{table}{0}
\setcounter{page}{1}
\makeatletter
\renewcommand{\theequation}{S\arabic{equation}}
\renewcommand{\thefigure}{S\arabic{figure}}
\renewcommand{\bibnumfmt}[1]{[S#1]}

\subsection*{Supplementary Information for ``Nonreciprocal quantum information processing with superconducting diodes in circuit quantum electrodynamics"}

\subsection*{S1. Superconducting Diode Effect from an asymmetric SQUID Junction} 
\label{AppendixA}
Following \cite{schrade2024dissipationless, souto2022josephson, fominov2022asymmetric}, we assume a superconducting quantum interference device (SQUID) with asymmetric junction potentials that will lead to a Josephson Diode effect. The diode behavior stems from the simultaneous breaking of inversion symmetry breaking and time reversal symmetry breaking. In contrast to bulk superconductors, where the effect stems from Cooper pais formed across different Fermi surfaces, in JJ based methods the effect occurs through imperfect transmission or anomalous phase accumulation across the junction \cite{nadeem2023superconducting}. The nonreciprocal behavior of the diode will manifests itself in the resonator structure through the kinetic inductance of the resonator. The bias of the diode is fixed via $\mathcal{T}$-breaking agent through $\Phi_b$ and $c_3$ which stems from the third order expansion of the Josephson potential and the signal across the ports sets the notion of directionality through the diode. 

\begin{figure}[h]
    \centering
    \includegraphics[width=0.33\linewidth]{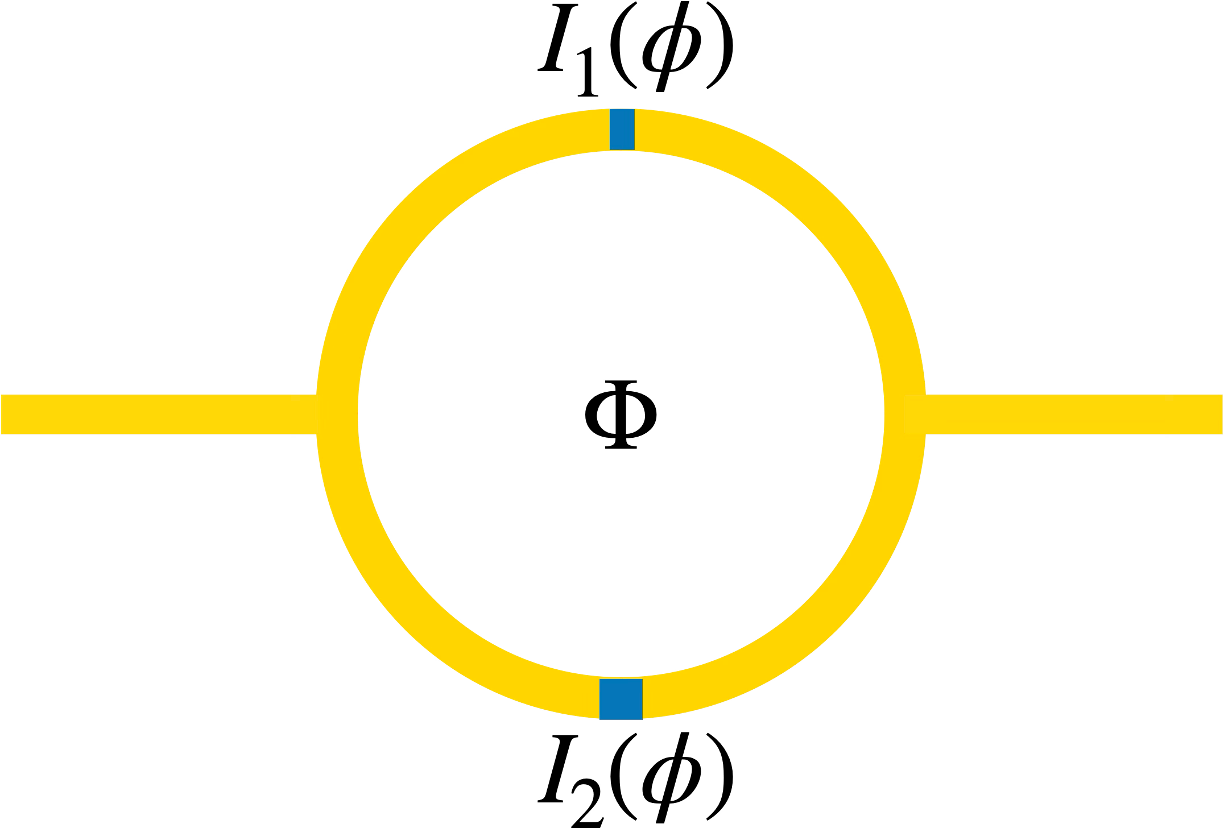}
    \caption{Asymmetric SQUID with different CPRs $I_1 (\phi)$ and $I_2 (\phi)$ in the two JJs.}
    \label{fig:SQUID}
\end{figure}

The the system Hamiltonian comprised of the diode coupler modeled as a capacitively shunted flux-tunable transmon comprised of two JJ's, transmission lines and interaction between them. Starting with the coupler systems we have the following Hamiltonian:
\begin{eqnarray}
\label{eq:resonator}
    \frac{H_{r}}{\hbar} = 4 E_c \Hat{n} + U(\Phi)
\end{eqnarray}
where $E_c = \frac{e^2}{2C}$ is the charge energy, $\Bar{n}$ the charge operator and $U(\Phi)=U_1(\Phi) + U_2(\Phi) $ represents the Josephson potential. The SQUID is pictured in \autoref{fig:SQUID} showcasing a simple JD interferometer setup comprising of two JJs (blue) with a nonsinusoidal current-phase relationship (CPR). The SQUID is flux controlled via $\Phi$. The diode effect stems from a simple model where two nonsinusoidal CPR's from the asymmetric junctions allow for higher order effects which are the crucial ingredient for nonreciprocity:

\begin{eqnarray}
    I_l (\Phi) &= \frac{e \Delta^2 \tau_l}{2\hbar} \frac{\sin(\Phi)}{\epsilon_l(\Phi)} \tanh\frac{\epsilon_l(\Phi)}{2k_BT} \\
    \epsilon_l(\Phi) &= \Delta \sqrt{1 - \tau_l \sin^2\frac{\Phi}{2}} .
\end{eqnarray}

The Josephson energy of each junction follows from CPR

\begin{eqnarray}
    U_l(\Phi) = -\Delta \sqrt{1 - \tau_l \sin^2\frac{\Phi}{2}}
\end{eqnarray}
In the SQUID with the two junction arrays the Josephson potential with an external flux $\Phi$ and a bias flux $\Phi_b$ can be readily defined as:
\begin{eqnarray}
    U(\Phi) = U_1(\Phi) + U_2(\Phi - \Phi_b)
\end{eqnarray}
This model allows us to describe the SD effect. 

\subsection*{S2. Transmission coefficient from classical input-output theory}
We implement classical input-output theory on a two-port network such that input and output on two ports are described by $\{a_1,a_2\}$ and $\{b_1,b_2\}$. Assuming input at port 1 and output at port 2, the transmission coefficient is written as
\begin{eqnarray}
    S_{21} = \frac{b_2}{a_1} = \frac{Z_0}{Z_0 + Z(\omega)}
\end{eqnarray}
where $Z(\omega) = R + i(\omega L - \frac{1}{\omega C})$. This leads us to define:
\begin{eqnarray}
    |S_{21}|^2 = \left|\frac{Z_0}{Z_0 + R + i(\omega L^\pm - \frac{1}{\omega C}})\right|^2
\end{eqnarray}
and after defining $\omega_0^\pm = \frac{1}{\sqrt{L^\pm C}}$ we obtain:
\begin{eqnarray}
    S_{21} = \frac{Z_0^2}{\frac{(Z_0 + R)^2}{4L^{\pm^2}} + (\omega - \omega_0^\pm)^2} \frac{1}{4L^{\pm^2}}
\end{eqnarray}
where $A = \frac{Z_0^2}{4L^2}$ and $\kappa = \frac{(Z_0 + R)^2}{4L^2}$ such that:
\begin{eqnarray}
    S_{21} = \frac{A^\pm}{\kappa^\pm + (\omega - \omega_0^\pm)^2}.
\end{eqnarray}

\subsection*{S3. Diode Spectroscopy}
Here we detail the spectroscopy calculations using equations of motion and quantum Hamiltonian. Combining the system Hamiltonian and analyzing the system with the transmission line, where the two ends of the transmission lines are coupled with rate $\kappa_1$ and $\kappa_2$, we can proceed and define the equations of motion coming from the Heisenberg-Langevin equations:
\begin{equation}
    \begin{aligned}
        \dot{a} = -\frac{i}{\hbar} [H_{r},a] - \frac{\kappa}{2} a + \sqrt{\kappa_1} a_{in} + \sqrt{\kappa_2} b_{in}
        \\ 
        a_{out} = a_{in} - \sqrt{\kappa_1} a
        \\
        b_{out} = b_{in} - \sqrt{\kappa_2} a
    \end{aligned}
\end{equation}
where $\kappa = \kappa_1 + \kappa_2$ represents the total decay. $a_{in} (a_{out})$ represents the input (output) ports of the system. The system is driven and probed with the standard Hamiltonian framework via $H_D =  (\epsilon_p e^{-i\omega_p t} a^\dagger - h.c) + (\epsilon_{pr} e^{-i\omega_{pr} t} a^\dagger - h.c.)  $  with drives $\epsilon_{p} = i\sqrt{\kappa} \epsilon_p$ and $\epsilon_{pr} = i\sqrt{\kappa} \epsilon_{pr}$ as well as frequencies $\omega_p$ and $\omega_{pr}$. 
We work in the rotating frame at the pump rate and also linearize around small quantum fluctuations such that $a(t) = \alpha e^{-i\omega_p t} + \delta a(t)$, where $\alpha$ is the classical amplitude coming from the coherent pump and $\delta a$ the fluctuations. One can then divide the equations of motion in two parts: one for the steady state amplitude of the classical coherent field and one for the fluctuations:
\begin{equation}
    \begin{aligned}
        \dot{\alpha} = -i (\Delta + \delta \omega_r + \frac{\Lambda}{2} |\alpha|^2  - i\frac{\kappa}{2}) \alpha  + \epsilon_p
        \\
        \dot{\delta a} (t) = -i(\Delta_{eff} - i\frac{\kappa}{2})\delta a - i \lambda \delta a^\dagger + \epsilon_{pr} e^{-i \Omega t}
    \end{aligned}
\end{equation}
where $\Delta_{eff} = \Delta + \delta \omega_r $, $\Delta = \omega_0 - \omega_p$ and $\lambda =  \Lambda |\alpha|^2$. With $\Lambda\neq 0$ we have included the Kerr nonlinearity for completeness. We apply the Fourier transform $\delta a (t) = \int d\Omega e^{-i\Omega t} d[\Omega]$ with $\Omega = \omega_{pr} - \omega_p$ the shifted probe frequency
\begin{eqnarray}
    \begin{pmatrix}
        i\Omega  + \frac{\kappa}{2} + i\Delta_{eff} & i\lambda \\
        -i\lambda^* & i\Omega  + \frac{\kappa}{2} - i\Delta_{eff}
    \end{pmatrix} \begin{pmatrix}
        \delta a[\Omega] \\
        \delta a^\dagger [-\Omega]
    \end{pmatrix}
    = \begin{pmatrix}
        \sqrt{\kappa_1} a_{in}[\Omega] + \sqrt{\kappa_2} b_{in}[\Omega]  \\
        \sqrt{\kappa_1} a_{in}^\dagger [-\Omega] + \sqrt{\kappa_2} b_{in}^\dagger [-\Omega] 
    \end{pmatrix}
\end{eqnarray}
where we define $M[\Omega]$ as the system matrix, where the boundary conditions remain unchanged with the linearized term $a = \alpha + \delta a$. We now invert $M[\Omega]$ to obtain the scattering parameters:
\begin{eqnarray}
    \chi(\Omega) = (i\Omega I - M[\Omega])^{-1} = \frac{1}{D(\Omega)} 
    \begin{pmatrix}
        i\Omega  + \frac{\kappa}{2} - i\Delta_{eff} & -i\lambda^* \\
        i\lambda & i\Omega  + \frac{\kappa}{2} + i\Delta_{eff}
    \end{pmatrix} , \\ \nonumber
    D(\Omega) = (i\Omega + \frac{\kappa}{2})^2 + \Delta_{eff}^2 - |\lambda|^2.
\end{eqnarray}
Now using the input-output relations and injecting an idler signal at $\omega = \Omega + \omega_p$ through port 1 such that $a_{in}[\Omega] = A_{in}$ and $b_{in}[\Omega] = 0$ we obtain the forward transmission spectrum. The fluctuation at $\omega$ becomes:
\begin{eqnarray}
    \delta a [\Omega] = \chi_{11} (\Omega) \sqrt{\kappa_1} A_{in} [\Omega] + \chi_{12}(\Omega) \sqrt{\kappa_1} A_{in} [-\Omega]
\end{eqnarray}
where $\chi(\Omega)_{ii} = (i\Omega I - M)^{-1}_{ii} (\Omega)$ is the susceptibility element from the linear response. The forward transmission coefficient then becomes:
\begin{eqnarray}
    S_{21} (\Omega) = \frac{b_{out} [\Omega]}{a_{in} [\Omega]} = -\sqrt{\kappa_1 \kappa_2} \chi_{11}(\Omega) -\sqrt{\kappa_1 \kappa_2} \chi_{12} (\Omega) \frac{A_{in} [-\Omega]}{A_{in} [\Omega]} 
\end{eqnarray}
where for backward transmission, we repeat the same method but instead inject the probe tone through $b_{in}$.

\subsection*{S4. Quantum analysis of diode nonreciprocity}
Starting from Eq.~(\ref{eq:resonator}), we expand the Josephson potential into Fourier components around $\Phi_{\min}$
\begin{eqnarray}
    U(\Phi) = \sum_n\frac{c_n}{n!} (\Phi-\Phi_{\rm min})^n,
\end{eqnarray}
and denote the expansion coefficients $c_n = \partial_{\Phi}^nU(\Phi)$, and $\Phi_{\rm min}$ is self-consistently found by solving $c_1(\Phi) = 0$. We focus on the term $H_3 = (c_3/6)\Phi^3$, and write flux in terms of bosonic operators
\begin{equation}
\label{eq:bosonoperatorSI}
    \Phi(x) = \sum_k \Phi_{\rm zpf} \left[\psi_k(x) a_k + \psi^*_k(x)a_k^\dagger\right]
\end{equation}
where $a_k$ is the bosonic operator, and $\psi_k(x) = u_k e^{ikx}$ is the mode function between the ports; $\Phi_{\rm zpf}$ is the flux at zero point fluctuation. We decompose flux into mean-field value and corresponding quantum fluctuations 
\begin{eqnarray}
   \Phi = \Phi_{b}+ \delta\Phi 
\end{eqnarray}
satisfying $\langle\Phi\rangle = \Phi_b$ and $\langle\delta\Phi\rangle = 0$. For bosonic operators this implies, $a_k = \alpha_k +d_k$ where $\langle a_k\rangle = \Phi_b\alpha_k$ is the mean-field expectation value satisfying $\langle\Phi\rangle = \Phi_b$ and $\langle d_k\rangle = 0$. With these definitions we can write 
\begin{equation}
    H_3 = \frac{c_3}{6} \prod_i \left(\sum_{k_i} \psi_{k_i} a_{k_i} \psi^*_{k_i} a_{k_i}^\dagger \right)
\end{equation}
and represent flux as
\begin{equation}
    \Phi = \mathcal{C} +X = \left(\sum_k \Phi_{\rm zpf} \left[\psi_k \alpha_k + \psi_k^* \alpha_k^*\right] \right) + \left(\sum_k \Phi_{\rm zpf} \left[\psi_k d_k + \psi_k^* d_k^*\right] \right)
\end{equation}
which together help us write
\begin{equation}
    H_3 = \frac{c_3}{6}\left[\mathcal{C}^3 + 3\mathcal{C}^2 X + 3 \mathcal{C}X^2 + X^3\right]
\end{equation}
where $\mathcal{C}\neq 0$ captures $\Phi_b\neq 0$. Note that for $\Phi_b =0$, the $X^3$ term given the three-wave mixing contribution, and for $\Phi_b \neq 0$ $3\mathcal{C}X^2$ gives the leading order bilinear correction which can shift the frequency of Hamiltonian in Eq.~(\ref{eq:resonator}). Moving on, we assume $C \propto \Phi_b$ as a constant and express $H_3$ as
\begin{equation}
    H_3 = \frac{\mathcal{C}c_3}{2} \Phi_{\rm zpf}^3\sum_{k k'} \psi_k \psi_{k'} d_k^\dagger d_{k'} = \sum_{k k'} \Lambda_{k k'} d_k^\dagger d_{k'}.
\end{equation}
We now narrow our discussion to a two mode system such as $a\equiv a_k$ is the forward moving mode and $b\equiv a_{-k}$ is the backward moving mode. We can then write the Hamiltonian as 
\begin{equation} 
    \tilde{H} = \begin{pmatrix}
        a^\dagger & b^\dagger
    \end{pmatrix}
    \begin{pmatrix}
        \omega_k + \Lambda_{k,k} & \Lambda_{k,-k} \\
        \Lambda_{-k,k} & \omega_{-k} + \Lambda_{-k,-k} 
    \end{pmatrix}
    \begin{pmatrix}
        a\\
        b
    \end{pmatrix},
\end{equation}
and it can be seen that both frequency shift and mode mixing arise from the interplay of $\Phi_b,c_3\neq 0$. However, mode mixing effects scale as $\Lambda^2$ and are thus even with $\Phi_b$. On the contrary, frequency shift is leading order in $\Lambda$ and thus emerge as tunable nonreciprocal signature. The corresponding difference in frequency of forward and backward mode then reads
\begin{equation}
    \delta\omega_r = \omega_k - \omega_{-k} = \frac{\mathcal{C}c_3}{2} \Phi_{\rm zpf}^3 \left[|u_k|^2 - |u_{-k}|^2\right].
\end{equation}
Noting that $\mathcal{C}\propto \Phi_b$, $c_3(\Phi_b) = -c_3(-\Phi_b)$ and $\left[|u_k|^2 - |u_{-k}|^2\right]$ flips sign with $\Phi_b$, we deduce that $\omega_k - \omega_{-k}$ is odd in $\Phi_b$, and reproduce Eq.~(\ref{eq:frshift}) of the main text.

\subsection*{S5. Flux Control of Diode Phase}
Starting from Eq. \ref{eq:qubitH} and \ref{eq:EOM}, we implement the flux control on the diode phase by incorporating different pulse shapes onto the diode phase into the effective qubit Hamiltonian \ref{eq:qubitH}  and the subsequently evolve the system through the established equations of motions \ref{eq:EOM}. Two pulse shapes were analyzed, namely a cosine pulse and a trapezoid pulse. Importantly, the trapezoid pulse can be engineered to allow for rectangular or triangular pulses to be implemented through simple variations of the control pulse definitions. 

\begin{figure}[h]
    \centering
    \includegraphics[width=0.75\linewidth]{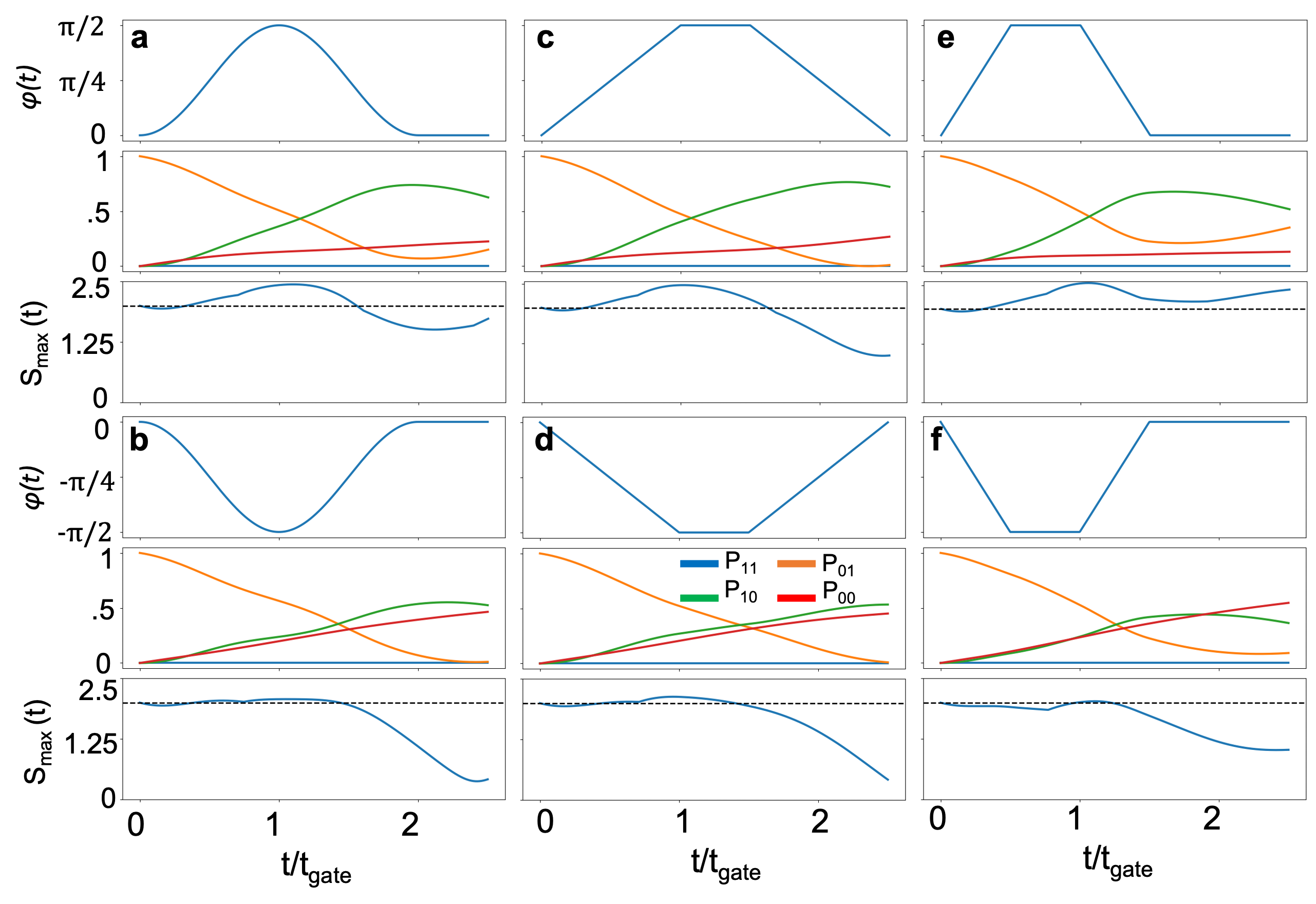}
    \caption{Pulse level control of the diode phase for different control pulses. (a-b) Cosine pulse applied to the diode phase up to maximum of $\varphi_{max} \pm \pi/2$. (c-d) Trapezoid pulse with longer ramp time $t_{hold} = \pi/(4J)$ and shorter hold time $t_{hold} = \pi/(8J)$ showcases worse entanglement dynamics than Fig. 4 (c-d) \ref{fig:BellStateTomo} in the main text. (e-f) Showcase trapezoid pulse with shorter hold time $t_{hold} = pi/(8J)$ but equal ramp up time as in Fig. \ref{fig:BellStateTomo}. The entanglement dynamics improve for $\varphi_{max} = \pi/2$, where the entanglement remains above the classical threshold $S_{max}$ for extended periods of the applied gate time $t_{gate}$. All plots were made with following parameters $\gamma_{1} = \gamma_{2} = \gamma_{\phi_1} = \gamma_{\phi_2} = 0$, $J=1$ unless mentioned otherwise.}
    \label{fig:control}
\end{figure}

The pulses are defined as:

\begin{align}
    \varphi(t) = 
    \begin{cases}
        0.0 & \textit{,if $t<0.0$} \\
        \varphi_{max} * f(\frac{t}{t_{ramp}}) & \textit{,if $t<t_{ramp}$} \\
        \varphi_{max} & \textit{,if $t<t_{ramp}+t_{hold}$} \\
        \varphi_{max} * (1-f(t-(t_{ramp} + t_{hold}))/t_{hold}) & \textit{,if $t<2*t_{ramp} + t_{hold}$}
    \end{cases}
\end{align}

where $f(t)$ defines the function specific to each pulse where for the cosine pulse we have $f(t) = cos(t)$ and for the trapezoid pulse we have $f(t) = t$. Fig. \ref{fig:control} displays the effects of different pulse shapes on the entanglement dynamics. The initial state for all evolutions start with $\ket{\Psi(t=0)} = \ket{01}$. Fig. \ref{fig:control} (a-b) show a cosine pulse for $\varphi_{max} \pm \pi/2$, where the entanglement dynamics and subsequent CHSH evolution follows a similar trajectory as in Fig. \ref{fig:BellStateTomo} in the main text. The smoother transition in the ramp up/down time leads to $S_{max}$ to die down for $\varphi_{max}$ after the pulse is turned off. Fig. \ref{fig:control} (c-f) show trapezoid pulses for various lengths in the ramp up/down time and hold time for the pulse sequence. Specifically in Fig. \ref{fig:control} (c-d), the hold time $t_{hold}$ at $\varphi_{max} \pm \pi/2$ was shortened to $\pi/(8J)$. The entanglement dynamics return quicker to their reciprocal setting once the ramp-down of the pulse occurs and remain below the $S_{max}=2$ threshold once the pulse is turned off. Fig. \ref{fig:control} (e-f) remain with the shorter hold time as previously but also the ramp up/down $t_{ramp}$ were equally shortened to $\pi/(8J)$. Here, due to the shorter length of the pulse, the nonreciprocal dynamics for $\varphi_{max} = \pi/2$ become starker such that once the pulse is turned off, the system remains above the $S_{max}=2$ threshold once the reciprocal dynamics settle back in, while for $\varphi_{max}=-\pi/2$, $S_{max}$ remains below 2 once the pulse is turned off and the reciprocal dynamics settle back in, demonstrating the nonreciprocity of our two qubit gate.

\subsection{S6. Full Bell-State Tomography}
\begin{figure}[h]
    \centering
    \includegraphics[width=1.0\linewidth]{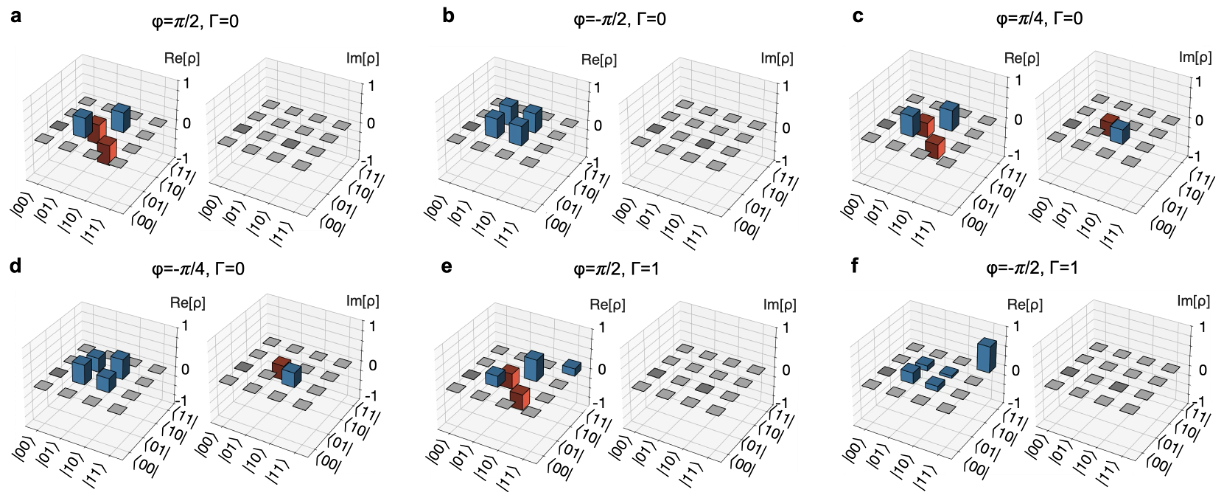}
    \caption{Bell state tomographic representation of density matrices of the qubit pairs via linear reconstruction after half-iSWAP ($t = \pi/4J$) applied to qubits. The system is initialized to $\Psi(t=0) = |01\rangle$ for all panels. (a-d) $\textbf{Re}[\rho]$ and $\textbf{Im}[\rho]$ without account of decays, $\gamma_1,\Gamma = 0$ for 
    $\varphi \in [\pi/2,-\pi/2, \pi/4, -\pi/4]$ reproducing Bell state derived in Eq.~(\ref{eq:bell-state-diode}). 
    Blue colors represent positive values while red represent negative values. 
    Tunable Bell-states are formed with non-trivial phase captured in $\textbf{Im}[\rho]\neq 0$.
    (e-f) Tomographic reconstruction of density matrix with collective cross decay, $\Gamma/J = 1$ at $\varphi = \pm \pi/2$. The Bell state generation is nonreciprocal determined by cooperative action of diode nonreciprocity and decay. For $\varphi = \pi/2$, $|\Psi_-\rangle$ is generated with nearly $80\%$ fideltiy and for $\varphi = -\pi/2$ no Bell state is formed as fidelity remains below 50\% (nearly 30\%). 
    This demonstrates the diodes ability to generate and distribute entanglement directionally between the qubit systems. 
    }
    \label{fig:FullBellTomo}
\end{figure} 

To verify that the nonreciprocal transient quantum correlation really corresponds to a Bell‐type superposition shown in Eq.~\ref{eq:bell-state-diode}, we perform full two‐qubit state tomography at the half-iSWAP time \(t = \pi/4J\). 
While our starting point is the numerically evaluated Bell state in presence of qubit relaxation ($\gamma_1$) and collective cross-decay ($\Gamma$), the tomographic representation provides benchmark signals to probe SD mediated nonreciprocal entanglement generation.  We begin by setting $\gamma_1 = \Gamma = 0$ to understand the effect of $\varphi\neq 0$ in Bell state in Eq.~(\ref{eq:bell-state-diode}) in terms of tunable entanglement generation. This directly reflects in real ($\textbf{Re}[\rho]$) and imaginary ($\textbf{Im}[\rho]$) parts of off-diagonal elements of density matrix. In Fig.~\ref{fig:FullBellTomo}a,b, we show the density matrix corresponding to $|\Psi(t=\pi/4J, \varphi = \pm\pi/2)\rangle$ with $|\Psi(t=0)\rangle = |10\rangle$ generating the usual Bell states $|\Psi_\mp\rangle = \frac{1}{2}(|10\rangle \mp |01\rangle)$, and $\textbf{Im}[\rho]=0$ throughout. The Bell state is flipped when $|\Psi(t=0)\rangle = |01\rangle$, and when $\varphi = 0$, same Bell pair $|\Psi(t=\pi/4J, \varphi = 0) = \frac{1}{2}(|10\rangle +i |01\rangle)$ is generated irrespective of system initialization. Similarly, other nontrivial $\varphi\neq 0$ can be tracked with Bell state reconstruction. For instance, $\varphi = \pm\pi/4$ is reflected as complex values of off-diagonal elements of reconstructed density matrix, see Fig.~\ref{fig:FullBellTomo}c,d. Next, we introduce the cross-qubit decay, $\Gamma\neq 0$ which facilitates passive SD nonreciprocity in isolation of generated Bell pairs. In Fig.~\ref{fig:FullBellTomo}e,f we show reconstructed density matrix for $\varphi = \pm\pi/2$, and we clearly note nonreciprocal entanglement generation. For simulated reconstructed density matrix, we find formation of $|\Psi_-\rangle$ for $\varphi = \pi/2$ with nearly $80\%$ fidelity for $\Gamma=J$. On the contrary, for $\varphi = \pi/2$, we find that fidelity remains below $50\%$ (nearly $30\%$) for $\Gamma = J$ and targeted entangled state is not generated. 
\end{document}